\documentclass[twocolumn,preprintnumbers,amsmath,aps]{revtex4}
\usepackage{graphicx}
\usepackage{dcolumn}
\usepackage{bm}
\usepackage{subfigure}
\usepackage[usenames]{color}
\begin{document}
%%%%%%%%%%%%%%%%%%%%%%%%%%%%
\newcommand{\kvec}{\mbox{{\scriptsize {\bf k}}}}
%%%%%%%%%%%%%%%%%%%%%%%%%%%%
\def\eq#1{(\ref{#1})}
\def\fig#1{\ref{#1}}
\def\tab#1{\ref{#1}}
%%%%%%%%%%%%%%%%%%%%%%%%%%%%
\title{Thermodynamics of the hydrogen dominant potassium hydride superconductor at high pressure}
\author{D. Szcz{\c{e}}{\'s}niak$^{1}$}\email{dszczesniak@qf.org.qa}
\author{R. Szcz{\c{e}}{\'s}niak$^{2, 3}$}
%%%%%%%%%%%%
\affiliation{1. Qatar Environment and Energy Research Institute, Qatar Foundation, PO Box 5825, Doha, Qatar}
\affiliation{2. Institute of Physics, Cz{\c{e}}stochowa University of Technology, Al. Armii Krajowej 19, 42-200 Cz{\c{e}}stochowa, Poland}
\affiliation{3. Institute of Physics, Jan D{\l}ugosz University in Cz{\c{e}}stochowa, 
Al. Armii Krajowej 13/15, 42-200 Cz{\c{e}}stochowa, Poland}
%%%%%%%%%%%%
\date{\today} 
\begin{abstract}
%%%%%%%%%%%%%%%%%%%%%%%%%%%%%%%%%%%%%%%%%%%%%%%%%%%
In the present paper we report comprehensive analysis of the thermodynamic properties of novel hydrogen dominant potassium hydride superconductor (KH$_{6}$). Our computations are conducted within the Eliashberg theory which yields quantitative estimations of the most important thermodynamic properties of superconducting phase. In particular, we observe, that together with the increasing pressure all the thermodynamic properties decrease, {\it e.g.} $T_{C} \in \left< 72.91 , 55.50 \right>$ K for $p \in \left< 166 , 300 \right>$ GPa. It is predicted that such decreasing behavior corresponds to the decreasing hydrogen lattice molecularization with increasing pressure value. Futhermore, by calculating the dimensionless thermodynamic ratios, familiar in the Bardeen-Cooper-Schrieffer theory (BCS), it is proved that KH$_{6}$ material is a strong-coupling superconductor and cannot be quantitatively described within the BCS theory.
%%%%%%%%%%%%%%%%%%%%%%%%%%%%%%%%%%%%%%%%%%%%%%%%%%%
\end{abstract}
\maketitle

%%%%%%%%%%%%%%%%%%%%%%%%%%%%%%%%%%%%%%%%%%%%%%%%%%%
\section{Introduction}
%%%%%%%%%%%%%%%%%%%%%%%%%%%%%%%%%%%%%%%%%%%%%%%%%%%

In the terms of research on the phonon-mediated superconductivity, the hydrogen-based materials \cite{ashcroft1} experience at present the apogee of the scientific attention \cite{szczesniak1}. The advent of this high interest is marked by the theoretical work of Ashcroft \cite{ashcroft2} who predicted that metallic hydrogen, under the influence of high-pressure, can be a superconductor characterized by the high transition temperature ($T_{C}$). Such desirable character of the superconducting phase in hydrogen stems from its light atomic nuclei and the absence of the inner electron shells, resulting in the high value of the electron-phonon coupling constant ($\lambda$). However, the pressure value required to induce the superconducting phase in pure hydrogen is beyond any commercial applications.

In this context, the introduction of the heavier atoms into the crystal lattice allows to lower the metallization pressure, preserving at the same time the extraordinary superconducting properties of hydrogen. In this context, it is possible to supplement the high-temperature superconductors \cite{szczesniak2}, by the ones described within the well-established electron-phonon scenario \cite{bardeen1}, \cite{bardeen2}. The true testimonial of the importance of the hydrogen-based superconductors are recent experimental results for the hydrogen sulfide superconductor (H$_{2}$S), which report the extremely high transition temperature equal to $\sim 190$ K, obtained at the relatively moderate pressure value ($p$) \cite{drozdov}, \cite{cartlidge}. If reproducible, these results may constitute a significant breakthrough in the research on the room-temperature superconductivity.

One of the newest attempts in the research on the superconducting properties of the hydrogen-based compounds consider also application of the alkali metal dopants. The most recent theoretical approach concentrates on using potassium, leading to the thermodynamically stable hydrogen-rich layered material where each six hydrogens hold one potassium atom (the KH$_{6}$ compound within the $C2/c$ phase) \cite{zhou}. In particular, it is theoretically predicted that superconducting phase, in this compound, is stable from $\sim 166$ GPa up to around 300 GPa. Please note that the upper pressure value constitutes the accuracy limit for the projector-augmented plane-wave potentials \cite{kresse1} used within the VASP code \cite{kresse2} for geometry optimization and electronic calculations in \cite{zhou}. Below the metallization pressure, but above 70 GPa, the KH$_{6}$ crystallizes in the metallic $C2/m$ phase. In what follows, the synthesis pressure for KH$_{6}$ is expected to be much lower then in the case of the other recently predicted alkali metal hydrides, namely the LiH-family \cite{zurek}. This fact arises from the faster transfer of electrons to the hydrogen in the KH$_{6}$ when comparing to the LiH-hydrides, and it is an undisputed advantage of this compound.

In this spirit, present paper reports the analysis of the thermodynamic properties of the KH$_{6}$ superconductor, which should be of crucial importance and interest for further design of possible hydride superconductors. The calculations are conducted here for three selected pressure values (166 GPa, 230 GPa, and 300 GPa) which sample the entire known superconducting phase of the KH$_{6}$ compound. Due to the relatively high values of the electron-phonon coupling constant for all considered pressure values, computations are conducted within the strong-coupling generalization of the Bardeen-Cooper-Schrieffer theory (BCS), namely in the framework of the Eliashberg equations \cite{eliashberg}. This theoretical model allows us to provide the quantitative estimations of all of the most important thermodynamic properties of the superconducting phase such as critical temperature, band gap at the Fermi level, specific heat, thermodynamic critical field, and the electron effective mass. We note that our analysis is based on the isotropic Elishaberg equations, following the character of the Eliashberg function ($\alpha^{2}F(\omega)$) presented in \cite{zhou}, and adopted for calculations in the present work.

%%%%%%%%%%%%%%%%%%%%%%%%%%%%%%%%%%%%%%%%%%%%%%%%%%%
\section{Theoretical model and computational details}
%%%%%%%%%%%%%%%%%%%%%%%%%%%%%%%%%%%%%%%%%%%%%%%%%%%

In order to estimate all desirable thermodynamic properties of the discussed KH$_{6}$ superconductor we solve the Eliashberg equations on the imaginary axis and in the mixed representation by using the iterative method presented in \cite{szczesniak3}.

As already mentioned, our calculations based on the Eliashberg spectral functions calculated in \cite{zhou}, by using the QUANTUM ESPRESSO code \cite{giannozzi} with the norm-conserving pseudopotentials within the generalized gradient approximation. Such spectral functions yields the electron-phonon coupling constants equal to: $\lambda_{p=166 {\rm GPa}}=0.92$, $\lambda_{p=230 {\rm GPa}}=0.89$, and $\lambda_{p=300 {\rm GPa}}=0.79$. Futhermore, we assume that the electron deparing correlations are described in the terms of the Coulomb pseudopotential ($\mu$) equals to 0.1 for all considered cases, and that the cutoff frequency is set as $\omega_{c}=10\Omega_{\rm max}$, where $\Omega_{\rm max}$ stands for the maximum phonon frequency equal to: 407.76 meV at $166$ GPa and 411.92 meV at $230$ GPa and $300$ GPa.

To assure the required precision, our computations are conducted for the 2201 Matsubara frequencies: $\omega_{m}\equiv\frac{\pi}{\beta}(2m-1)$, with $\beta\equiv 1/k_{B}T$, and $k_{B}$ denoting the Boltzmann constant. In what follows, all thermodynamic properties of interest are described quantitatively for $T\geq T_{0}\equiv 5$ K.

%%%%%%%%%%%%%%%%%%%%%%%%%%%%%%%%%%%%%%%%%%%%%%%%%%%
\section{Numerical results}
%%%%%%%%%%%%%%%%%%%%%%%%%%%%%%%%%%%%%%%%%%%%%%%%%%%

\begin{figure}[ht!]
\includegraphics[width=\columnwidth]{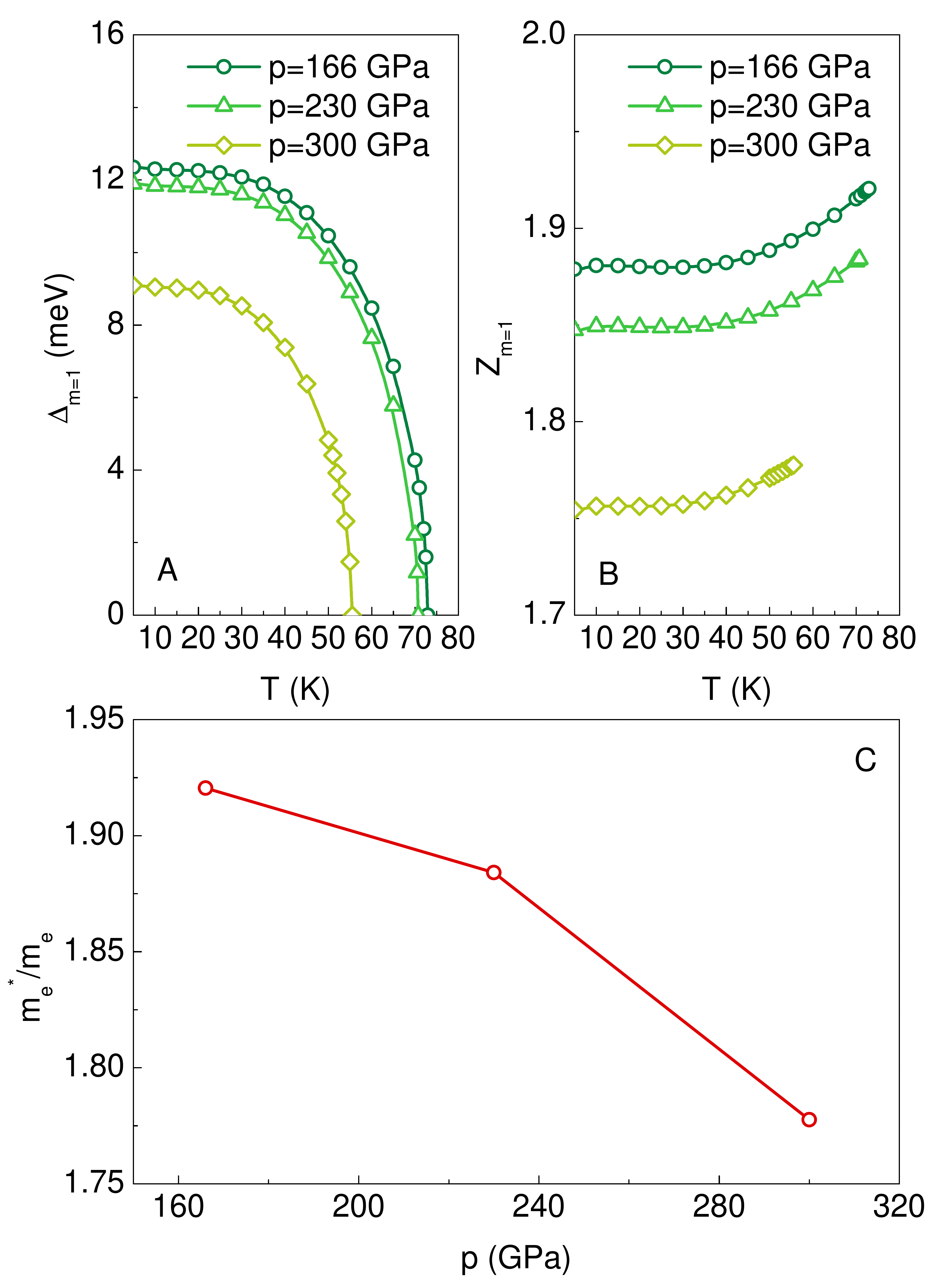}
\caption{The maximum values of the order parameter (A), wave function renormalization factor (B) and the normalized electron effective mass (C). First two parameters are presented as a function of temperature for selected values of pressure. The latter observable is plotted against the pressure value.}
\label{fig01}
\end{figure}

In Fig. \ref{fig01} we present obtained imaginary-axis results of the (A) maximum value of the order parameter ($\Delta_{m=1}$), (B) the wave function renormalization factor ($Z_{m=1}$), and (C) the normalized electron effective mass ($m_{e}^{*}/m_{e}$, where $m^{\star}_{e}\simeq Z_{m=1} m_{e}$ is the electron effective mass, and $m_{e}$ denotes the bare electron mass). 

In particular, for all considered pressure cases we observe decrease in the values of the $\Delta_{m=1}$, $Z_{m=1}$, and $m_{e}^{*}/m_{e}$ functions with the increasing pressure. In what follows, such a decrease is also observed in the terms of the superconducting transition temperature calculated by using the relation: $\Delta_{m=1}(T_{C})=0$, which gives the $T_C$ values equal to $72.91$ K, $70.73$ K, and $55.50$ K at $166$ GPa, $230$ GPa, and $300$ GPa, respectively. We also note, that this decreasing behavior of $T_{C}$ is slightly more evident in the case of our results then in the predictions presented in \cite{zhou}, which based on the McMillan formula \cite{mcmillan2}. Furthermore, results presented in \cite{zhou} show that for higher electron-phonon coupling constants McMillan formula underestimate the $T_{C}$ values in KH$_{6}$, whereas in the case when $\lambda$ is close to the weak-coupling limit ($\lambda < 0.3$ \cite{cyrot}) the overestimation can be noticed. This is due to the fact that BCS theory omits the strong-coupling effects, which are present in the Eliashberg theory.

\begin{figure}[ht]
\includegraphics[width=\columnwidth]{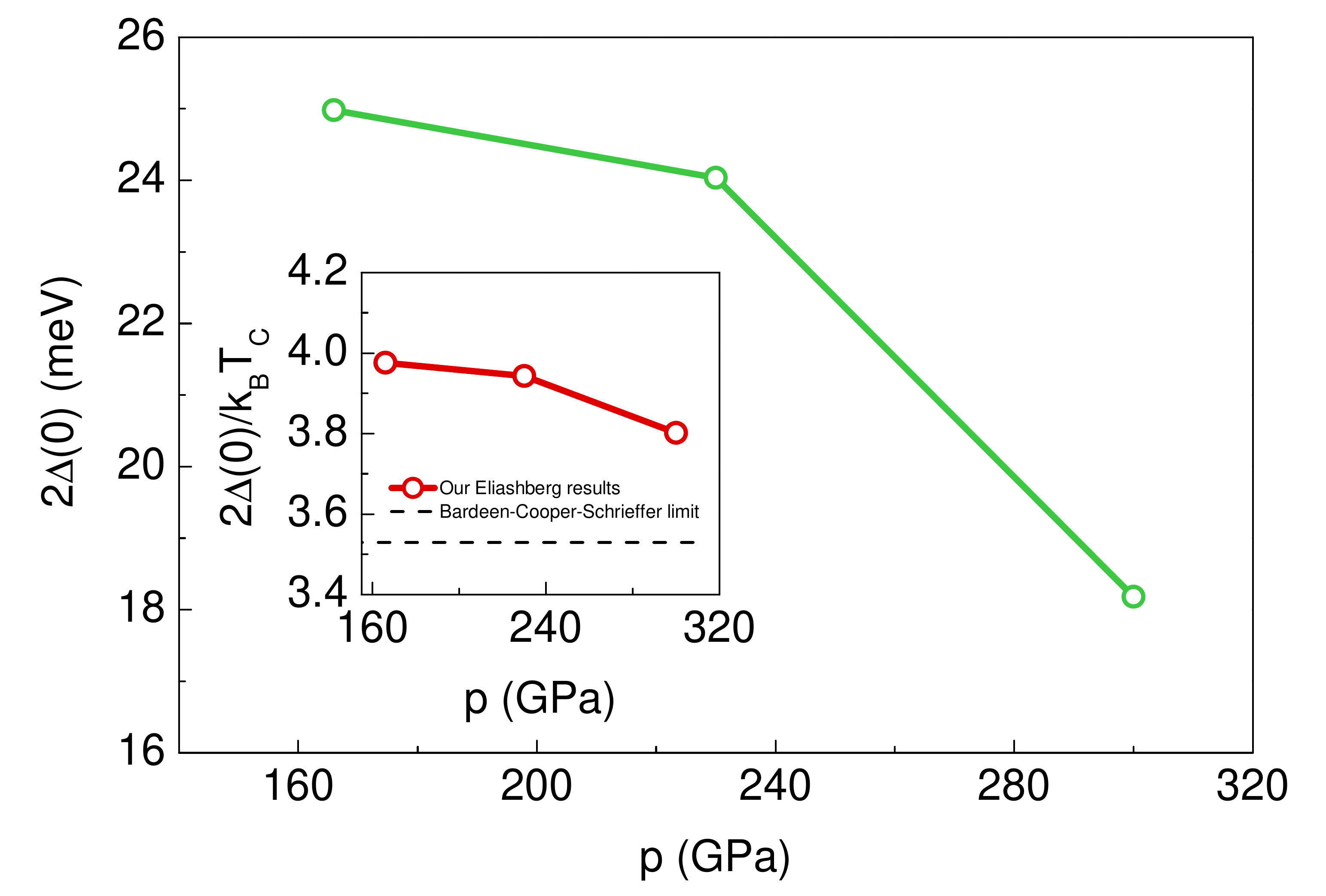}
\caption{The physical value of the energy gap at the Fermi energy ($2\Delta(0)$) as a function of pressure. The inset depicts pressure dependance of the $2\Delta\left(0\right)/k_{B}T_{C}$ characteristic dimensionless ratio, which is compared with the predictions of the BCS theory.}
\label{fig02}
\end{figure}

The corresponding results for the value of energy gap at the Fermi level are presented in Fig. \ref{fig02} as a function of pressure. We note that, in the first approximation such calculations can be done on the basis of the imaginary-axis results presented in Fig. \ref{fig01}, as $2\Delta_{m=1} (0)$, where $\Delta_{m=1}\left(0\right)\simeq\Delta_{m=1}\left(T_{0}\right)$. However, in order to obtain the physical value of the energy gap ($2\Delta(0)$, where $\Delta(0)\simeq\Delta(T_0)$) it is required to analytically continue the imaginary-axis results on the real axis ($\omega$). In this context the physical value of the energy gap can be determined by using $\Delta\left(T\right)={\rm Re}\left[\Delta\left(\omega=\Delta\left(T\right),T\right)\right]$.

Furthermore, in the inset of Fig. \ref{fig02} we depict determined values of the characteristic dimensionless ratio of zero-temperature energy gap at the fermi level to the critical temperature ($R_{\Delta}\equiv 2\Delta\left(0\right)/k_{B}T_{C}$). We note that for all considered pressure values the $R_{\Delta}$ ratios notably exceed limit set by the BCS theory ($R_{\Delta}^{\rm BCS}=3.53$), \cite{bardeen1}, \cite{bardeen2}. In particular, $R_{\Delta}$ equals to 3.98, 3.94, and 3.80 for 166 GPa, 230 GPa, and 300 GPa, respectively.

\begin{figure}[ht]
\includegraphics[width=\columnwidth]{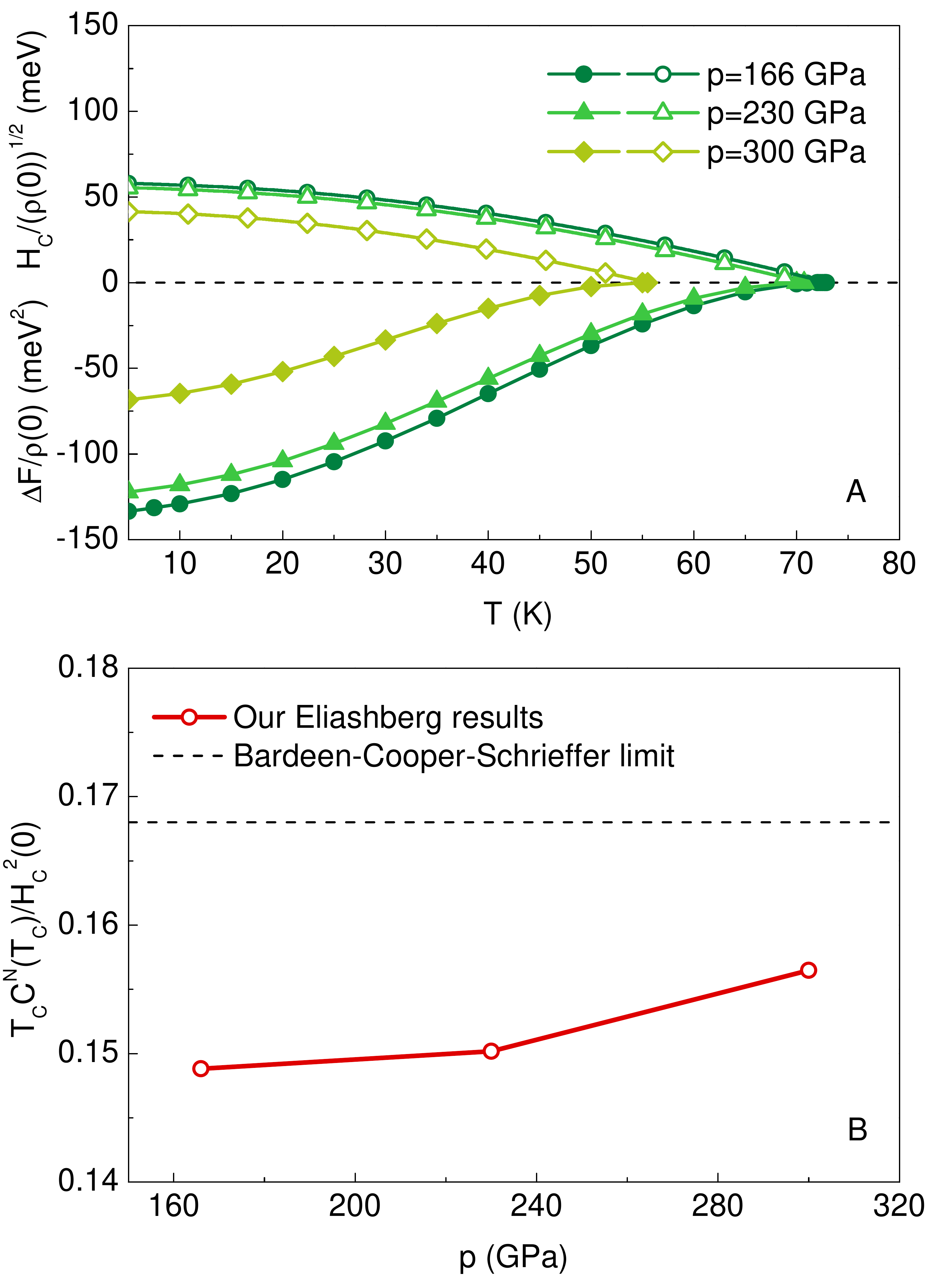}
\caption{(A) The normalized free energy ($\Delta F / \rho\left(0\right)$) and thermodynamic critical field ($H_{C} / \sqrt{\rho\left(0\right)}$) as a function of temperature, for selected pressure values. (B) The $T_{C}C^{N}\left(T_{C}\right)/H_{C}^{2}\left(0\right)$ characteristic dimensionless ratio as a function of pressure, along with the predictions of the BCS theory marked by horizontal line.}
\label{fig03}
\end{figure}

In the next step, our calculations concentrate on the determination of the normalized free energy difference between the superconducting and normal state ($\Delta F / \rho\left(0\right)$, where $\rho(0)$ is the electron density of states at the Fermi level). For this purpose, we use the following relation:
\begin{eqnarray}
\label{r10}
\frac{\Delta F}{\rho\left(0\right)}&=&-\frac{2\pi}{\beta}\sum_{n=1}^{M}
\left(\sqrt{\omega^{2}_{n}+\Delta^{2}_{n}}- \left|\omega_{n}\right|\right)\\ \nonumber
&\times&(Z^{S}_{n}-Z^{N}_{n}\frac{\left|\omega_{n}\right|}
{\sqrt{\omega^{2}_{n}+\Delta^{2}_{n}}}),
\end{eqnarray}  
where $Z^{S}_{n}$ and $Z^{N}_{n}$ represent the wave function renormalization factors for the superconducting ($S$) and normal ($N$) state, respectively. Obtained numerical results are presented in the lower panel of Fig. \ref{fig03} (A) for all considered pressure values. We note, that the $\Delta F / \rho\left(0\right)$ function takes negative values for $T \in \left<T_{0}, T_{C} \right>$. This fact ensures the thermodynamic stability of the superconducting phase in the KH$_{6}$ material.

By using the determined $\Delta F / \rho\left(0\right)$ function we are next able to estimate the value of normalized thermodynamic critical field ($H_{C} / \sqrt{\rho\left(0\right)}$) as:

\begin{equation}
\label{r11}
\frac{H_{C}}{\sqrt{\rho\left(0\right)}}=\sqrt{-8\pi\left[\Delta F/\rho\left(0\right)\right]}.
\end{equation}

The corresponding results are presented in the upper panel of the Fig. \ref{fig03} (A) as a function of temperature. Again calculations are conducted for the three considered pressure values. For both $\Delta F / \rho\left(0\right)$ and $H_{C} / \sqrt{\rho\left(0\right)}$ parameters their absolute value decreases together with the increasing pressure. The inverse behavior is observed for the dimensionless ratio which corresponds to the zero-temperature thermodynamic critical field ($R_{H}\equiv T_{C}C^{N}\left(T_{C}\right)/H_{C}^{2}\left(0\right)$, where $H_{C}^{2}\left(0\right)\simeq H_{C}^{2}\left(T_{0}\right)$), as presented in Fig. \ref{fig03} (B). Moreover, similarly like in the case of the $R_{\Delta}$ parameter, the values of the $R_{H}$ ratio differ notably from the BCS predictions ($R_{H}^{\rm BCS}=0.168$), \cite{bardeen1}, \cite{bardeen2}. In particular, $R_{H}=0.149$, $R_{H}=0.150$, and $R_{H}=0.156$, for 166 GPa, 230 GPa, and 300 GPa, respectively.

\begin{figure}[ht]
\includegraphics[width=\columnwidth]{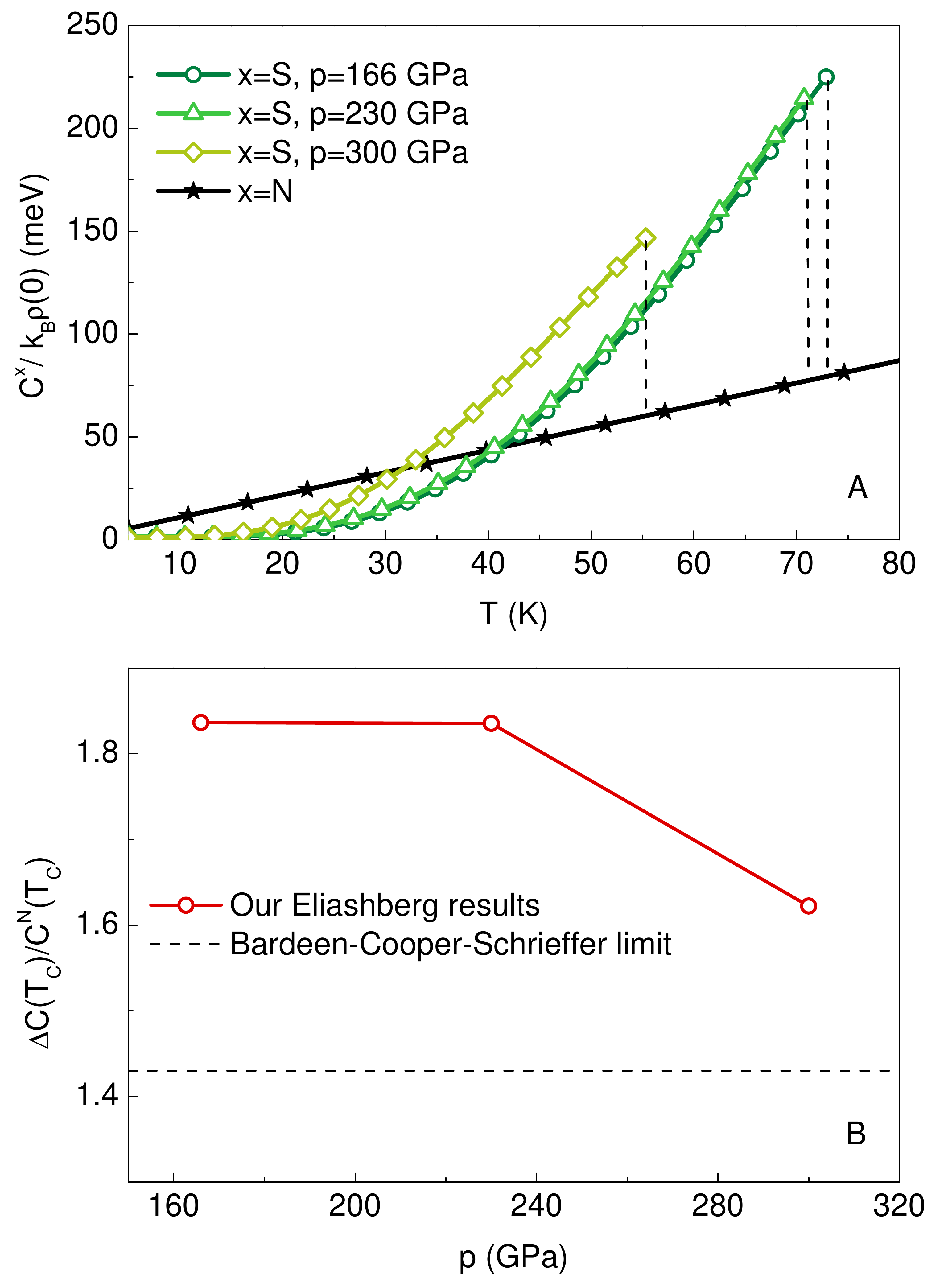}
\caption{(A) The specific heat of the superconducting ($C^{S}$) and normal ($C^{N}$) state as a function of temperature for three considered pressure values. (B) The $\Delta C\left(T_{C}\right) / C^{N}\left(T_{C}\right)$ characteristic dimensionless ratio as a function of pressure. The BCS predictions are marked by horizontal line.}
\label{fig04}
\end{figure}

In the last step, we analyze the temperature behavior of the specific heat for the superconducting ($C^{S}$) and normal ($C^{N}$) state. In what follows, the specific heat of the normal state is calculated as:
\begin{equation}
\label{r13}
\frac{C^{N}\left(T\right)}{ k_{B}\rho\left(0\right)}=\frac{\gamma}{\beta}, 
\end{equation}
where $\gamma\equiv\frac{2}{3}\pi^{2}\left(1+\lambda\right)$ denotes the Sommerfeld constant. In this manner the specific heat of the superconducting state ($C^{S}$) can be obtained by using the following relation:
\begin{equation}
\label{r12}
C^{S}=C^{N}+\Delta C,
\end{equation}
where $\Delta C$ is the difference between the specific heat of the normal and the superconducting state, which can be calculated as:
\begin{equation}
\label{r12}
\frac{\Delta C\left(T\right)}{k_{B}\rho\left(0\right)}=-\frac{1}{\beta}\frac{d^{2}\left[\Delta F/\rho\left(0\right)\right]}{d\left(k_{B}T\right)^{2}}.
\end{equation}
The results for $C^{N}$ and $C^{S}$ functions are presented in Fig. \ref{fig04} (A). For all considered pressure values, we observe the characteristic specific heat jump at the critical temperature, which is marked by the vertical line. Moreover, like in the case of previously calculated thermodynamic properties the $C^{S}$ function decreases with the increasing pressure. Finally, determined dimensionless ratio for the specific heat ($R_{C}\equiv \Delta C\left(T_{C}\right) / C^{N}\left(T_{C}\right)$) equals $R_{C}=1.84$ for 166 GPa and 230 GPa, and $R_{C}=1.62$ for 300 GPa, respectively, as presented in Fig. \ref{fig04} (B). Again obtained results exceed limit set by the BCS theory ($R_{C}^{\rm BCS}=1.43$), \cite{bardeen1}, \cite{bardeen2}.

%%%%%%%%%%%%%%%%%%%%%%%%%%%%%%%%%%%%%%%%%%%%%%%%%%%
\section{Summary}
%%%%%%%%%%%%%%%%%%%%%%%%%%%%%%%%%%%%%%%%%%%%%%%%%%%

In this paper, by using the Eliashberg formalism, all the most important thermodynamic properties are determined for the hydrogen dominant KH$_{6}$ superconductor at high-pressure. In particular, we obtain the quantitative estimations of the parameters such as critical temperature, order parameter, energy gap at the Fermi level, thermodynamic critical field, and the specific heat. All results are given for the three considered pressure values, which sample the entire known KH$_{6}$ superconducting phase ($p \in \left< 166 {\rm GPa}, 300 {\rm GPa} \right>$).

On the basis of our analysis we observe decrease in the values of the calculated thermodynamic properties with increasing pressure. This fact can be followed on the example of the critical temperature value which drops from 72.91 K to 55.50 K when pressure is raised from 166 GPa to 300 GPa. From the physical point of view, such decreasing behavior corresponds to the decreasing hydrogen molecularization and stronger electron depairing correlations with increasing pressure.

Our results are additionally summarized by the calculations of the characteristic dimensionless ratios, familiar in the BCS theory, which correspond to the zero-temperature energy gap at the Fermi level ($R_{\Delta}$), the zero-temperature thermodynamic critical field ($R_{H}$), and the specific heat for the superconducting state ($R_{C}$). We note, the determined values of the $R_{\Delta} \in \left< 3.98, 3.80 \right>$, $R_{H} \in \left< 0.149, 0.156 \right>$, and $R_{C} \in \left< 1.84, 1.62 \right>$ ratios at $p \in \left< 166 , 300 \right>$ GPa notably differ from the predictions of the BCS theory for the weak-coupling limit ($R_{\Delta}^{\rm BCS}=3.53$, $R_{H}^{\rm BCS}=0.168$, and $R_{C}^{\rm BCS}=1.43$). These differences arise due to the strong-coupling and retardation effects which occur in the KH$_{6}$ superconductor and proves that this material cannot be quantitatively analyzed within the BCS theory which lacks the proper description of such effects.
  
%%%%%%%%%%%%%%%%%%%%%%%%%%%%%%%%%%%%%%%%%%%%%%%%%%%
\begin{acknowledgments}

We are grateful to the Cz{\c{e}}stochowa University of Technology - MSK CzestMAN for granting access to the computing infrastructure built in the project No. POIG.02.03.00-00-028/08 "PLATON - Science Services Platform".
\end{acknowledgments}
%%%%%%%%%%%%%%%%%%%%%%%%%%%%%%%%%%%%%%%%%%%%%%%%%%%
\bibliographystyle{apsrev}
\bibliography{manuscript}
%%%%%%%%%%%%%%%%%%%%%%%%%%%%%%%%%%%%%%%%%%%%%%%%%%%
\end{document}